\documentclass[preprint,12pt]{elsarticle}

\usepackage{bm}
\usepackage{amsmath,amssymb,amsfonts,times}
\usepackage{url}

\usepackage[colorlinks=true, pdfstartview=FitV, linkcolor=red, citecolor=blue, urlcolor=blue]{hyperref}

\usepackage{graphicx}
\usepackage{epsfig}
\usepackage{slashed}
\usepackage{a4wide}
\usepackage{fancyhdr}
\usepackage{subfigure}

\usepackage{rotating}

\usepackage{pdflscape}

\allowdisplaybreaks[1]

\begin{document}

\newcommand{\UNIT}[1]{\mbox{$\,{\rm #1}$}}
\newcommand{\MeV}{\UNIT{MeV}}
\newcommand{\GeV}{\UNIT{GeV}}
\newcommand{\GeVc}{\UNIT{GeV/c}}
\newcommand{\TeV}{\UNIT{TeV}}
\newcommand{\AMeV}{\UNIT{AMeV}}
\newcommand{\AGeV}{\UNIT{AGeV}}
\newcommand{\ATeV}{\UNIT{ATeV}}
\newcommand{\fm}{\UNIT{fm}}
\newcommand{\mb}{\UNIT{mb}}
\newcommand{\fmc}{\UNIT{fm/c}}
\newcommand{\proz}{\UNIT{\%}}
\newcommand{\ds}{\displaystyle}
\newcommand{\zerovec}{\vec{0}\,}
\newcommand{\qvec}{\vec{q}\,}
\newcommand{\kvec}{\vec{k}\,}
\newcommand{\lvec}{\vec{l}\,}
\newcommand{\nvec}{\vec{n}\,}
\newcommand{\pvec}{\vec{p}\,}
\newcommand{\rvec}{\vec{r}\,}
\newcommand{\jvec}{\vec{j}\,}
\newcommand{\Gcapvec}{\vec{G}\,}
\newcommand{\Pcapvec}{\vec{P}\,}
\newcommand{\Picapvec}{\vec{\Pi}\,}
\newcommand{\Scapvec}{\vec{S}\,}
\newcommand{\Tcapvec}{\vec{T}\,}
\newcommand{\lambdavec}{\vec{\lambda}\,}
\newcommand{\sigmavec}{\vec{\sigma}\,}
\newcommand{\omegavec}{\vec{\omega}\,}
\newcommand{\sigmacapvec}{\vec{\Sigma}\,}
\newcommand{\thetacapvec}{\vec{\Theta}\,}
\newcommand{\tauvec}{\vec{\tau}\,}
\newcommand{\rhovec}{\vec{\rho}\,}
\newcommand{\deltavec}{\vec{\delta}\,}
\newcommand{\pivec}{\vec{\pi}\,}
\newcommand{\xivec}{\vec{\xi}\,}
\newcommand{\xvec}{\vec{x}\,}
\newcommand{\nrmsq}{\mbox{$\langle r^2\rangle_n$}}
\newcommand{\be}{\begin{equation}}
\newcommand{\ee}{\end{equation}}
\newcommand{\ba}{\begin{eqnarray}}
\newcommand{\ea}{\end{eqnarray}}
\newcommand{\etal}{\mbox{\it et al.}}
\newcommand{\prim}{\hspace{-1mm}'}
\def\bm#1{\mbox{\boldmath$#1$} }
\newcommand{\rp}{\textsc{r/p}-}
\newcommand{\derivl}{\stackrel{\leftarrow}{\partial}}
\newcommand{\derivr}{\stackrel{\rightarrow}{\partial}}

\newcommand{\evl}{e^{\frac{i\stackrel{\leftarrow}{\partial}_{\beta}v^{\beta}+m}{\Lambda}}}
\newcommand{\evr}{e^{\frac{-v^{\beta}i\stackrel{\rightarrow}{\partial}_{\beta}+m}{\Lambda}}}
\newcommand{\esl}{e^{\frac{i\stackrel{\leftarrow}{\partial}_{\beta}v^{\beta}+m}{\Lambda}}}
\newcommand{\esr}{e^{\frac{-v^{\beta}i\stackrel{\rightarrow}{\partial}_{\beta}+m}{\Lambda}}}

\newcommand{\err}{e^{\frac{-v^{\beta}i\stackrel{\rightarrow}{\partial}_{\beta}+m}{\Lambda}}}
\newcommand{\Gv}{\frac{g_{\omega}}{\Lambda}}
\newcommand{\Gr}{\frac{g_{\rho}}{\Lambda}}
\newcommand{\Gs}{\frac{g_{\sigma}}{\Lambda}}

\newcommand{\evnml}{e^{\frac{E}{\Lambda}}}
\newcommand{\esnml}{e^{\frac{E}{\Lambda}}}

\newcommand{\evnm}{e^{-\frac{E-m}{\Lambda}}}
\newcommand{\esnm}{e^{-\frac{E+m}{\Lambda}}}

\newcommand{\Gva}{\frac{g_{\omega}}{\Lambda}}
\newcommand{\Gsa}{\frac{g_{\sigma}}{\Lambda}}

\newcommand{\Gvb}{\frac{g_{\omega}}{ (\Lambda)^{2}}}
\newcommand{\Gsb}{\frac{g_{\sigma}}{ (\Lambda)^{2}}}

\newcommand{\fac}{\frac{\kappa}{(2\pi)^{3}}}

\newcommand{\nld}{{\cal D}}
\newcommand{\calo}{{\cal \Omega}}

\newcommand{\nldl}{\overleftarrow{{\cal D}}}
\newcommand{\nldr}{\overrightarrow{{\cal D}}}

\newcommand{\calol}{\overleftarrow{\varOmega}}
\newcommand{\calor}{\overrightarrow{\varOmega}}

\newcommand{\pspace}{\int\limits_{|\pvec|\leq p_{F_{i}}}\!\!\!\!\!\! d^{3}p}

\newcommand{\partialr}{\overrightarrow{\partial}}
\newcommand{\partiall}{\overleftarrow{\partial}}

\newcommand{\xil}{\overleftarrow{\xi}}
\newcommand{\xir}{\overrightarrow{\xi}}

\newcommand{\chil}{\overleftarrow{\chi}}
\newcommand{\chir}{\overrightarrow{\chi}}

\newcommand{\dleftp}{\nldl^{\prime}\big|_{\hat{\xi}=0} \xil^{\alpha}}

\newcommand{\drightp}{\nldr^{\prime}\bigg|_{\hat{\xi}=0}\xir^{\alpha}}

\newcommand{\dleftpp}{\nldl^{\prime\prime}\big|_{\hat{\xi}=0} \xil^{\alpha}\xil^{\beta}}

\newcommand{\drightpp}{\nldr^{\prime\prime}\bigg|_{\hat{\xi}=0}\xir^{\alpha}\xir^{\beta}}

%%% e-exp
\newcommand{\forma}{e^{-\zeta^{\alpha}i\partialr_{\alpha}+m/\Lambda}}
\newcommand{\formanm}{e^{-\frac{E-m}{\Lambda}}}
%%% p-monopole
\newcommand{\formb}{\frac{1}{1+\sum_{j=1}^{4}\left(\zeta_{j}^{\alpha} \, i\partialr_{\alpha}\right)^{2}}}
\newcommand{\formbnm}{\frac{\Lambda^2}{ \Lambda^2+\vec{p}^{\,2}}}
%%% p-dipole
\newcommand{\formc}{\left[\frac{1}{1+\sum_{j=1}^{3}\left(\zeta_{j}^{\alpha} \, i\partialr_{\alpha}\right)^{2}}\right]^{2}}
\newcommand{\formcnm}{\left[\frac{1}{ 1+\left(\frac{p}{\Lambda}\right)^{2}}\right]^{2}}

\newcommand{\NPA}{Nucl.~Phys.~}
\newcommand{\PR}{Phys.~Rep.~}   
\newcommand{\PL}{Phys.~Lett.~}
\newcommand{\PRC}{Phys.~Rev.~}
\newcommand{\PRL}{Phys.~Rev.~Lett.~}
\newcommand{\EPJ}{Eur.~Phys.~J.~}

\newcommand{\Bla}{\Big<}
\newcommand{\Bra}{\Big>}
%\maxdeadcycles=100000

\newcommand{\panda}{$\overline{\mbox P}$ANDA~}

%%%%%%%%%%%%%%%%%%%%%%%%%%%%%%%%%%%%%%%%%%%%%%%%%%%%%%%%%%%%%%%%%%
\begin{frontmatter}

\title{Production of multi-strangeness hypernuclei and the YN-interaction}

\author{T.~Gaitanos$^{1}$, H.~Lenske$^{1,2}$}
\address{$^{1}$ Institut f\"ur Theoretische Physik, Universit\"at Giessen,
             D-35392 Giessen, Germany}
\address{$^{2}$ GSI Helmholtzzentrum f\"ur Schwerionenforschung, D-64291 Darmstadt, Germany}
\address{email: Theodoros.Gaitanos@theo.physik.uni-giessen.de}

\begin{abstract}
We investigate for the first time the influence of hyperon-nucleon (YN) interaction 
models on the strangeness dynamics of antiproton- and $\Xi$-nucleus interactions. 
Of particular interest is the formation of bound 
multi-strangeness hypermatter in reactions relevant for \panda. The main features of two 
well-established microscopic approaches for YN-scattering are first discussed and their results 
are then analysed such that they can be applied in transport-theoretical simulations. The transport 
calculations for reactions induced by antiproton beams on a primary target including also the 
secondary cascade beams on a secondary target show a strong sensitivity on the underlying YN-interaction. 
In particular, we predict the formation of $\Xi$-hypernuclei with an observable 
sensitivity on the underlying $\Xi$N-interaction. 
We conclude the importance of our studies 
for the forthcoming research plans at FAIR.
\end{abstract}

\end{frontmatter}

\date{\today}

%%%%%%%%%%%%%%%%%%%%%%%%%%%%%%%%%%%%%%%%%%%%%%%%%%%%%%%%%%%%%%%%%%%%%%%%%%%%%%%
\section{\label{sec1}Introduction}
%%%%%%%%%%%%%%%%%%%%%%%%%%%%%%%%%%%%%%%%%%%%%%%%%%%%%%%%%%%%%%%%%%%%%%%%%%%%%%%

A central task of flavor nuclear physics is the construction of a realistic 
physical picture of nuclear forces between the octet-baryons $N,~\Lambda,~\Sigma,~\Xi$
~\cite{Haidenbauer,gal,Gibson:1995an,HypFirst1,SchaffnerBielich:2008kb}. 
Experimental information on the bare hyperon-nucleon interactions is accessible 
in the S=-1 sector ($\Lambda N$ and $\Sigma N$ 
channels)~\cite{S1a,S1b,Valcarce:2005em,S1d,S1e}, 
however, empirical data for the S=-2 channels involving the cascade hyperon
are still very sparse~\cite{S2data}. 
As a consequence, the parameters of the bare YN-interactions in the S=-1 channel are 
better under control than those parameters in the S=-2 sector, e.g., 
$\Xi N\to\Xi N,~\Lambda\Lambda$. In fact, 
various theoretical models, which are based on the well-established one-boson-exchange 
approach for the $NN$-interaction or rely on more sophisticated models on the quark 
level, predict quite different results for the 
$\Xi N$-channels in vacuum~\cite{S1d,S2data}. This is very clearly manifested in the different energy 
dependence of phase-shifts, which in one model are compatible with an attractive~\cite{rijken}, 
and in another approach with a repulsive~\cite{fuwi} $\Xi N$-interactions in free space. 
Of extreme interest is here the $\Xi N\to \Lambda\Lambda$ channel, because it provides 
information also on the bare interaction between hyperons themselves and it is the leading 
channel for the production of double-$\Lambda$ hypernuclei. 

Information of these YN interactions at finite baryon density can be achieved in studies 
of hypernuclei~\cite{Gaitanos:2013rxa,Steinheimer:2012gq,Botvina:2012zza} in 
reactions induced by mesons ($\pi$- and $K$-beams), 
by high-energy (anti)protons and heavy-ions, and by electro-production~\cite{ref15,ref16}. 
Recently, the FOPI~\cite{Herrmann:2012lea} and HypHI~\cite{HypHI} Collaborations at GSI 
have performed experiments on single-$\Lambda$ hypernuclei with the analysis being still in progress. 
The experimental investigation of multi-strangeness, e.g., double-$\Lambda$, hypernuclei 
will be realized in the new FAIR facility at GSI by the \panda 
Collaboration~\cite{panda1,panda2}. According to the \panda proposals, multi-strangeness 
bound hypermatter is supposed to be created in a two-step process through the capture of 
cascade particles ($\Xi$) - produced in primary antiproton-induced reactions - on 
secondary targets. Also theoretical investigations have been 
started recently~\cite{Gaitanos:2013rxa,Steinheimer:2012gq,Botvina:2012zza,Botvina:2007pd}.

We have studied in the past the formation and production mechanisms of 
hyperons~\cite{Larionov:2011fs}, fragments~\cite{Gaitanos:2007mm} 
and hyperfragments~\cite{Gaitanos:2013rxa,Gaitanos:2009at} in reactions 
induced by heavy-ions, protons and antiprotons, however, by using fixed interactions for 
the hyperon-nucleon channels, in particular, in the S=-2 sector. The task of the present work 
is to investigate the role and possible observable effects of different YN-interaction 
models on the strangeness dynamics of hadron-induced reactions. We have extended our earlier 
works by considering improved parametrizations for the cross sections of the S=-2 YN-channels. They 
are based on the microscopic calculations of the Nijmegen group by Rijken et al.~\cite{rijken} and 
by Fujiwara et al.~\cite{fuwi}. In sect.~\ref{sec2} we briefly outline the main features of the 
adopted microscopic YN-calculations and discuss the basic differences between them in terms of 
scattering observables. For numerical purposes, the theoretical cross sections have been 
parametrized and implemented 
into a transport model. We discuss then the transport results of 
hadron-induced reactions in detail. Our results show strong 
dynamical effects originating from the different underlying $\Xi N$-interactions. They 
are clearly visible in the yields of multi-strangeness hypernuclear production in 
low-energy $\Xi$-induced reactions. Hence, the proposed production experiments may lead 
to strong constraints on the high strangeness YN and YY interactions. 

%%%%%%%%%%%%%%%%%%%%%%%%%%%%%%%%%%%%%%%%%%%%%%%%%%%%%%%%%%%%%%%%%%%%%%%%%%%%%%%%%%%
\section{\label{sec2} Hyperon-nucleon interaction models in the S=-2 sector}
%%%%%%%%%%%%%%%%%%%%%%%%%%%%%%%%%%%%%%%%%%%%%%%%%%%%%%%%%%%%%%%%%%%%%%%%%%%%%%%%%%%

A variety of well-established models concerning the high strangeness sector exists in the 
literature. Among others, the chiral-unitary approach 
of Sasaki, Oset and Vacas~\cite{S1e} and the effective field theoretical models of the 
Bochum/J\"ulich groups~\cite{S1e} are representative examples. The Nijmegen~\cite{rijken} 
models are based on the well-known one-boson-exchange formalism to the NN-interaction, 
which is then extended to the strangeness sector with the help of SU(3) symmetry arguments. 
Finally, Fujiwara et al.~\cite{fuwi} have developed quark-cluster models for the 
baryon-baryon (BB) interactions in the S=0,-1,-2,-3,-4 sectors. 

%%%%%%%%%%%%%%%%%%
\begin{figure}[t]
\begin{center}
\includegraphics[clip=true,width=0.8\columnwidth,angle=0.]
{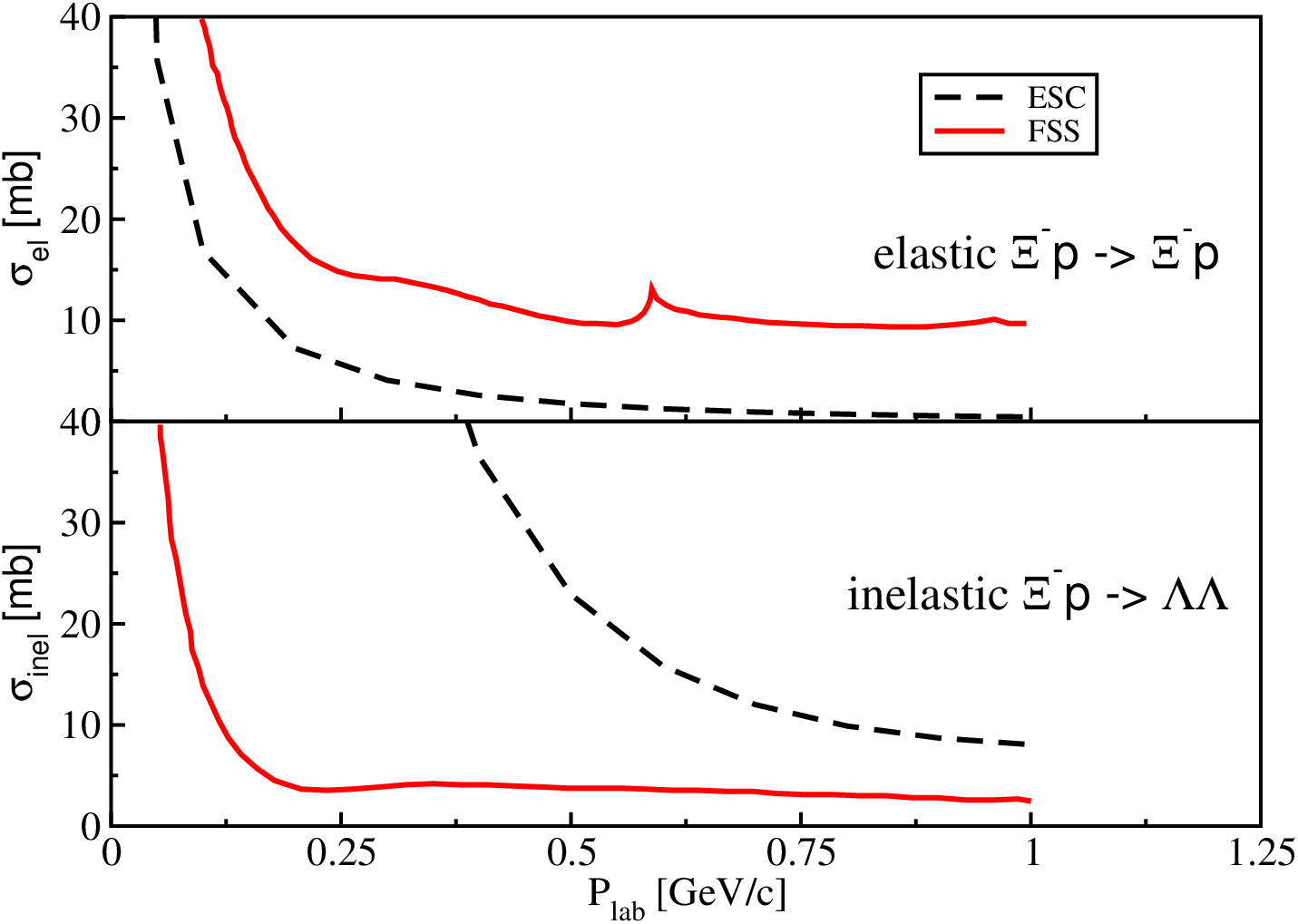}
\caption{\label{fig1}
Elastic (panel on the top) and inelastic (panel on the bottom) cross sections versus the 
$\Xi$-beam momentum for the $\Xi^{-}$p-channel. Results obtained by the ESC approach 
(dashed curves,~\cite{rijken}) are compared to the FSS model (solid curves,~\cite{fuwi}).
\vspace{-0.3cm}
}
\end{center}
\end{figure}
%\vspace{-1.0cm}
%%%%%%%%%%%%%%%%%%
The parameters are determined by simultaneous fits to NN- and YN-scattering 
observables in the S=-1 sector. While there are several thousand well confirmed 
experimental data in the S=0 sector (NN scattering), much less data points are safely known 
for YN reactions.
%While in the S=0 sector (NN-interactions) 
%there are about 4300 scattering data of high quality available, in the S=-1 YN-part only 
%38 scattering data are accessible. 
They still allow a reasonable determination of the 
$\Lambda$N and $\Sigma$N model parameters. 
Thus, despite of model differences all the theoretical approaches yield similar 
predictions for, e.g., potential depths and scattering cross sections for exclusive channels 
between nucleons and hyperons with strangeness S=-1. Remaining uncertainties for 
S=-1 interactions are related to the unsatisfactory experimental data base. 

In the S=-2 sector, where cascade particles ($\Xi$) are involved, the uncertainties and, 
correspondingly, the discrepancies 
between the models are considerably larger~\cite{S2data}. To demonstrate this issue, we have 
chosen two particular model calculations to be tested in the present work. 
The one-boson-exchange calculations of the Nijmegen group in the extended soft-core version ESC04 
of 2004 (we call this as ESC in the following)~~\cite{rijken} and the quark-cluster 
approach~\cite{fuwi} (denoted as FSS in the following). Note that more modern 
approaches of the Nijmegen group exist, e.g., the more recent models ESC08~\cite{S1d}. However, 
our intention is to investigate possible sensitivities in the dynamics of hadron-induced 
reactions at \panda, thus we choose two models which utilize very different physical pictures 
and exhibit the largest differences in the $\Xi N$-interactions. 

Fig.~\ref{fig1} shows the elastic and inelastic scattering cross sections 
for the isospin I=1 $\Xi N\to \Xi N$ and $\Xi N\to \Lambda\Lambda$ channels, respectively, 
predicted by the two models as indicated. We show cross sections for the relevant exclusive 
channels only, as entering into the dynamical calculations (see next section). 
At first, the ESC calculations predict a stronger 
energy dependence of the $\Xi N$ scattering. This effect is extreme at lower cascade energies, 
where the inelastic $\Xi N\to \Lambda\Lambda$ channel dominates relative to the elastic one. 
In fact, this pronounced energy dependence has been observed indirectly in the production of 
double-strangeness $\Lambda$-hypernuclei, where the double-$\Lambda$ hypernuclei yields dropped 
considerably with increasing $\Xi$-beam energy~\cite{Gaitanos:2013rxa}. On the other hand, the 
FSS calculations predict a much smoother energy behaviour of both cross 
sections, where an opposite trend relative to the ESC results is observed: the S=-2 dynamics 
is largely dominated by the $\Xi$N channel, while the $\Lambda\Lambda$ channel is populated to 
a lesser degree.

The differences in the cross sections reflect the rather different predictions of the scattering 
parameters, e.g., the phase shifts for low-energy s-states. The Nijmegen model 
%in the ESC04a 
in the ESC04 
version, which is used here, predict negative s-wave phase shifts for the I=0 $\Xi N$ channel 
at various cascade energies indicating a repulsive interaction, 
while in the $\Lambda\Lambda$ case the opposite trend is obtained. Indeed, as explicitly 
pointed out in~\cite{rijken}, no bound $\Xi N$ states are expected in the 
%ESC04a parameter 
ESC04 parameter 
set. On the other hand, the FSS model calculations~\cite{fuwi} exhibit a 
completely different energy behaviour of the phase-shifts for the same I=0 $\Xi N$ channel. 
The phase shift values are smaller and, in particular, are carrying positive signs indicating 
an attractive $\Xi N$-interaction. 

Note that both models lead also to different predictions for the single-particle (s.p.) in-medium 
potential of $\Lambda$ and, in particular, of the cascade particles. As discussed in detail 
in Refs.~\cite{S1d,kohno} in the framework of G-matrix calculations, the in-medium s.p. 
$\Lambda$-potential is in both cases attractive with 
values between -38\MeV~and -45\MeV~in the ESC and FFS models, respectively, for matter at saturation 
density following roughly the quark-scaling rule of a reduction by a factor of $2/3$ with respect to 
the NN-case. The in-medium cascade potentials are rather uncertain, ranging at saturation density from repulsion (ESC) to weak attraction (FFS). In order to keep the present discussion transparent, 
we will use for the s.p. mean-field potentials the same quark-counting scheme for both models. 

The question arises whether such different model predictions for the S=-2 YN-interaction 
can be constrained experimentally by \panda. In spite of the pronounced model dependencies, 
an answer on this question is not trivial, because of secondary processes. Indeed, 
in contrast to free space scattering, in hadron-induced nuclear reactions sequential 
re-scattering also occurs, for instance, quasi-elastic hyperon-nucleon re-scattering 
with strangeness exchange ($\Lambda N \leftrightarrow \Sigma N$). 

For this purpose we use the elastic and inelastic cross sections 
for $\Xi N$-scattering for both models in a newly derived parametrizations. 
In our approach we are accounting also for a smooth transition into 
the high-energy regime ($E>3.4$ \GeV~for baryon-baryon collisions), 
where the PYTHIA~\cite{pythia} generator is switched on. 
The parametrizations used for numerical purpose in the transport calculations, 
which fit the ESC and FSS calculations very well, have been extracted with 
piecewise polynomial fits.  

%%%%%%%%%%%%%%%%%%%%%%%%%%%%%%%%%%%%%%%%%%%%%%%%%%%%%%%%%%%%%%%%%%%%%%%%%%%%%%%%%%%
\section{\label{sec3}Transport-theoretical approach to multi-strangeness production}
%%%%%%%%%%%%%%%%%%%%%%%%%%%%%%%%%%%%%%%%%%%%%%%%%%%%%%%%%%%%%%%%%%%%%%%%%%%%%%%%%%%

For the theoretical description of the antiproton-induced primary reactions and the 
subsequently generated secondary reactions with a $\Xi$-beam we adopt the well-established 
relativistic Boltzmann-Uheling-Uhlenbeck (BUU) transport 
approach~\cite{Botermans:1990qi}.  
The kinetic equations are numerically realized within the Giessen-BUU (GiBUU) transport 
model~\cite{Buss:2011mx}, where the transport equation is given by
%%%%%%%%%%%%%%%
\begin{equation}
\left[
k^{*\mu} \partial_{\mu}^{x} + \left( k^{*}_{\nu} F^{\mu\nu}
+ m^{*} \partial_{x}^{\mu} m^{*}  \right)
\partial_{\mu}^{k^{*}}
\right] f(x,k^{*}) = {\cal I}_{coll}
%\frac{1}{2(2\pi)^9} \nonumber\\
%& & \times \int \frac{d^3 k_{2}}{E^{*}_{{\bf k}_{2}}}
%             \frac{d^3 k_{3}}{E^{*}_{{\bf k}_{3}}}
%             \frac{d^3 k_{4}}{E^{*}_{{\bf k}_{4}}} W(kk_2|k_3 k_4)
% \left[ f_3 f_4 \tilde{f}\tilde{f}_2 -f f_2 \tilde{f}_3\tilde{f}_4
%\right] 
\quad .
\label{rbuu}
\end{equation}
%%%%%%%%%%%%%%%
Eq.~(\ref{rbuu}) describes the dynamical evolution of the one-body phase-space 
distribution function $f(x,k^{*})$ for the hadrons under the influence of a 
hadronic mean-field (l.h.s. of Eq.~(\ref{rbuu})) and binary collisions (r.h.s. of Eq.~(\ref{rbuu})).
The mean-field is treated within the relativistic mean-field approximation 
of Quantum-Hadro-Dynamics~\cite{qhd}. It enters into the transport equation through 
the kinetic $4$-momenta $k^{*\mu}=k^{\mu}-\Sigma^{\mu}$ and effective (Dirac) 
masses $m^{*}=M-\Sigma_{s}$. The in-medium self-energies, 
$\Sigma^{\mu} = g_{\omega}\omega^{\mu} + \tau_{3}g_{\rho}\rho_{3}^{\mu}$ and 
$\Sigma_{s} = g_{\sigma}\sigma$, describe the in-medium interaction of nucleons 
($\tau_{3}=\pm 1$ for protons and neutrons, respectively). 
The isoscalar, scalar $\sigma$, the isoscalar, vector $\omega^{\mu}$ and the 
third isospin-component of the isovector, vector meson field $\rho_{3}^{\mu}$ 
are obtained from the standard Lagrangian equations of motion~\cite{qhd}. 
The obvious parameters (meson-nucleon couplings) are taken from the widely used 
parametrizations including non-linear self-interactions of the $\sigma$ field ~\cite{lala}. 
The meson-hyperon couplings at the mean-field level are obtained 
from the nucleonic sector using simply quark-counting arguments. The collision 
term includes all necessary binary processes for (anti)baryon-(anti)baryon, 
meson-baryon and meson-meson scattering and annihilation~\cite{Buss:2011mx}. 
Important for the present work is the implementation of the new parametrizations for 
the $\Xi$N-scattering processes, as discussed in the previous section. Having the 
cross sections for all relevant exclusive elementary channels, the collision integral 
of the transport equation is then modelled with standard Monte-Carlo techniques. 

%%%%%%%%%%%%%%%%%%%%%%%%%%%%%%%%%%%%%%%%%%%%%%%%%%%%%%%%%%%%%%%%%%%%%%%%%%%%%%%%%%%
\section{Strangeness production in $\bar{p}$-induced reactions}
%%%%%%%%%%%%%%%%%%%%%%%%%%%%%%%%%%%%%%%%%%%%%%%%%%%%%%%%%%%%%%%%%%%%%%%%%%%%%%%%%%%
%%%%%%%%%%%%%%%%%%
\begin{figure}[t]
%\vspace{-5cm}
\begin{center}
\includegraphics[clip=true,width=0.8\columnwidth,angle=0.]{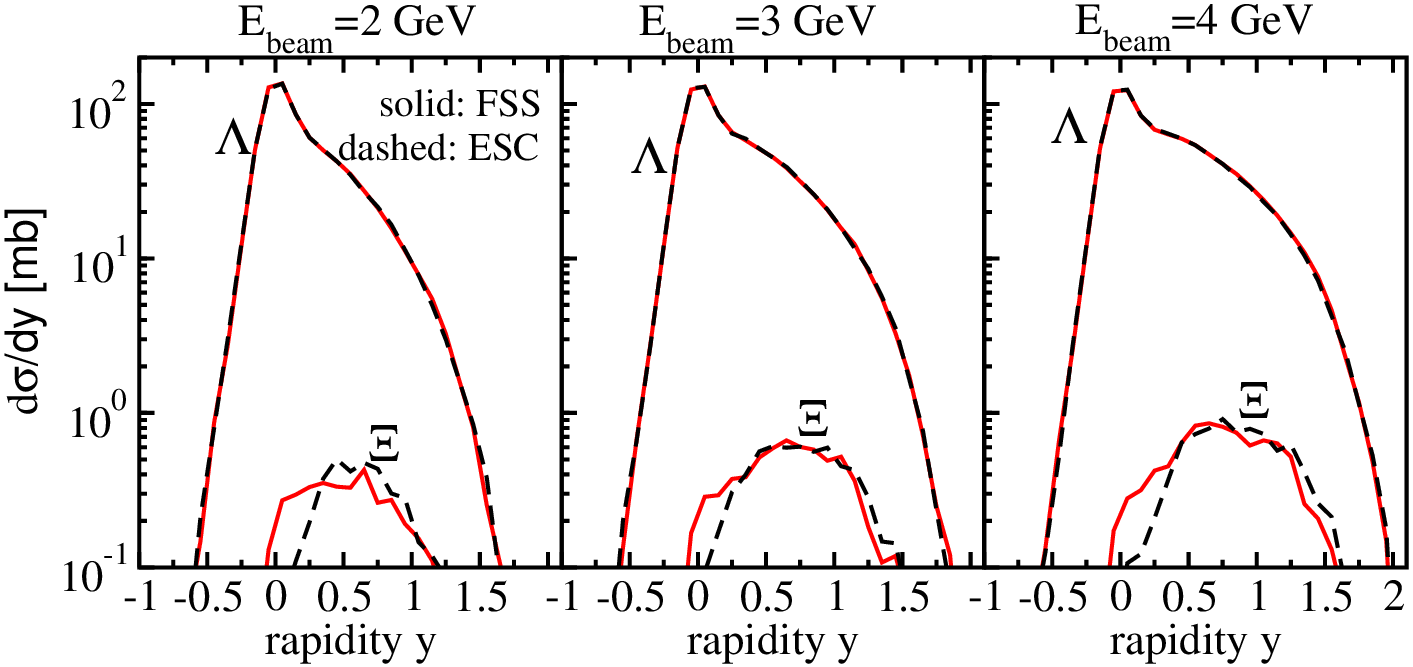}
%\vspace{-1.5cm}
\caption{\label{fig2}
Rapidity distributions of $\Lambda$ and $\Xi$ hyperons, as indicated, for $\bar{p}$-induced 
reactions on Cu-target at $\bar{p}$-beam energies of 2, 3 and 4 \GeV~ for the panel on the 
left, middle and right, respectively. Transport calculations with the $\Xi$N-parametrizations 
according to the ESC~\cite{rijken} (dashed curves) and to the FSS~\cite{fuwi} (solid curves) 
models are shown.
\vspace{-0.3cm}
}
\end{center}
\end{figure}
%%%%%%%%%%%%%%%%%%
We have performed transport calculations for $\bar{p}$-induced reactions including 
the secondary collisions of $\Xi$-beams and on a second target. We focus our studies 
on the role of the $\Xi$N-interaction models on the reaction dynamics and start the 
discussion with the primary $\bar{p}$-induced reactions. 

Fig.~\ref{fig2} show the rapidity spectra of $\Lambda$ and $\Xi$ hyperons at 
$\bar{p}$-beam energies of 2,3 and 4~\GeV~on a Cu target. 
At first, the S=-1 hyperon distributions are not affected by the choice of the $\Xi$N 
model. This is due to the fact that the $\Lambda$ hyperons are mainly produced 
by $\bar{p}p$ primary annihilation and, in particular, get re-distributed by many secondary 
processes involving sequential re-scattering of antikaons ($\bar{K}$) and 
hyperonic S=-1 resonances. The secondary processes in the S=-1 sector make the 
$\Lambda$ distributions broad and unaffected by the particular treatment of the $\Xi$N 
interaction. Hence, $\Lambda$ production will serve to explore independently the S=-1 sector.

The production of $\Xi$ particles is a comparatively rare process, as clearly seen in 
Fig.~\ref{fig2} by the strong decrease of the $\Xi$-rapidity distributions. This is due to the 
small values of the direct annihilation cross section $\bar{p}p\to\bar{\Xi}\Xi$, which 
is in the range of a few micro-barns only~\cite{landolt}. Note that the corresponding 
annihilation cross section into $\bar{\Lambda}\Lambda$ pairs is in the range of a few hundreds 
of micro-barn~\cite{landolt}. As in the case of $\Lambda$ particles, re-scattering contributes 
also to the broadness of the $\Xi$ rapidity spectra. However, the effect is less pronounced 
here due to the high production threshold of the $\Xi$ particles. They mainly escape from 
the residual excited target nucleus. 

%%%%%%%%%%%%%%%%%%
\begin{figure}[t]
%\vspace{-5cm}
\begin{center}
\includegraphics[clip=true,width=0.8\columnwidth,angle=0.]{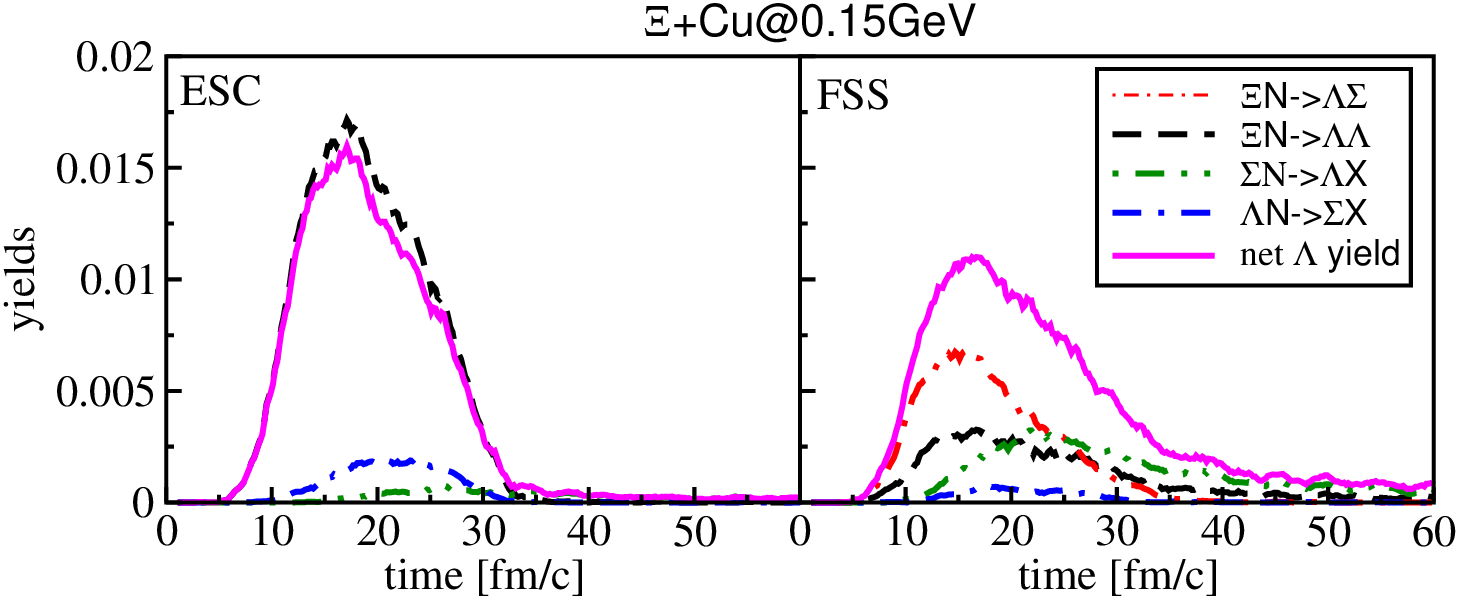}
%\vspace{-1.0cm}
\caption{\label{fig3}
Time evolution of the net $\Lambda$ yields (solid curves) and of the contributions 
of the exclusive channels $\Xi N\to \Lambda\Lambda$ (dashed curves), 
$\Xi N\to \Lambda\Sigma$ (dot-dashed curves), $\Sigma N\to \Lambda X$ (dot-dot-dashed curves) 
and $\Lambda N\to \Sigma X$ (dot-dashed-dashed curves), where $X$ stands for everything else. 
The yields are normalized to the number of events. Transport calculations using the ESC 
(left panel) and FSS (right panel) parametrizations are shown for central (b=0fm) 
$\Xi$-induced reaction on Cu-target at a beam $\Xi$-energy of 0.15 \GeV.
\vspace{-0.3cm}
}
\end{center}
\end{figure}
%%%%%%%%%%%%%%%%%%
The $\Xi$-yields are rather stable and only moderately dependent on the choice of the 
$\Xi$N-interaction model. The results 
with the FSS $\Xi$N-cross sections (solid curves in Fig.~\ref{fig2}) lead to increased 
re-scattering and thus to a shift of the $\Xi$-spectra to lower energies, relative to the 
transport calculations using the ESC $\Xi$N-cross sections (dashed curves in Fig.~\ref{fig2}). 
This is obvious by looking at the high-energy part of the elastic cross sections in 
Fig.~\ref{fig1}, where $\sigma_{el}$ strongly decreases with increasing $\Xi$-energy, 
in particular, in the ESC calculations. 

%%%%%%%%%%%%%%%%%%%%%%%%%%%%%%%%%%%%%%%%%%%%%%%%%%%%%%%%%%%%%%%%%%%%%%%%%%%%%%%%%%%
\section{Multi-Strangeness production in $\Xi$-induced reactions}
%%%%%%%%%%%%%%%%%%%%%%%%%%%%%%%%%%%%%%%%%%%%%%%%%%%%%%%%%%%%%%%%%%%%%%%%%%%%%%%%%%%

In \panda the low-energy part of the cascade particles is expected to be used in a second 
step as a secondary beam. They will react with a secondary target and may produce 
double-strangeness hypernuclei thus giving access to S=-2 hypermatter~\cite{panda1}. 
Since the underlying theoretical models 
for the $\Xi$N-interaction exhibit the largest differences at low energies, it is of 
great interest to study in detail the role of the different $\Xi$N-approaches to the dynamics 
of $\Xi$-induced reactions. In order to investigate such processes in this section, we consider 
the $\Xi$N-reactions as "primary" reactions, although under realistic experimental conditions 
the $\Xi$ hyperons are produced in an initial annihilation process. 

Fig.~\ref{fig3} shows the time evolution of the net $\Lambda$ yield (solid curves) together 
with the corresponding contributions to $\Lambda$ production/absorption, for $\Xi$-induced 
reactions on a Cu-target. At first, both sets of 
transport calculations result to a similar total net $\Lambda$ yield. The calculations using 
ESC (FSS) parametrizations lead to values of 0.242 (0.243), respectively, for the 
total net $\Lambda$ relative yield (normalized to the number of GiBUU events). 

%%%%%%%%%%%%%%%%%%
\begin{figure}[t]
%\vspace{-2cm}
\begin{center}
\includegraphics[clip=true,width=0.85\columnwidth,angle=0.]{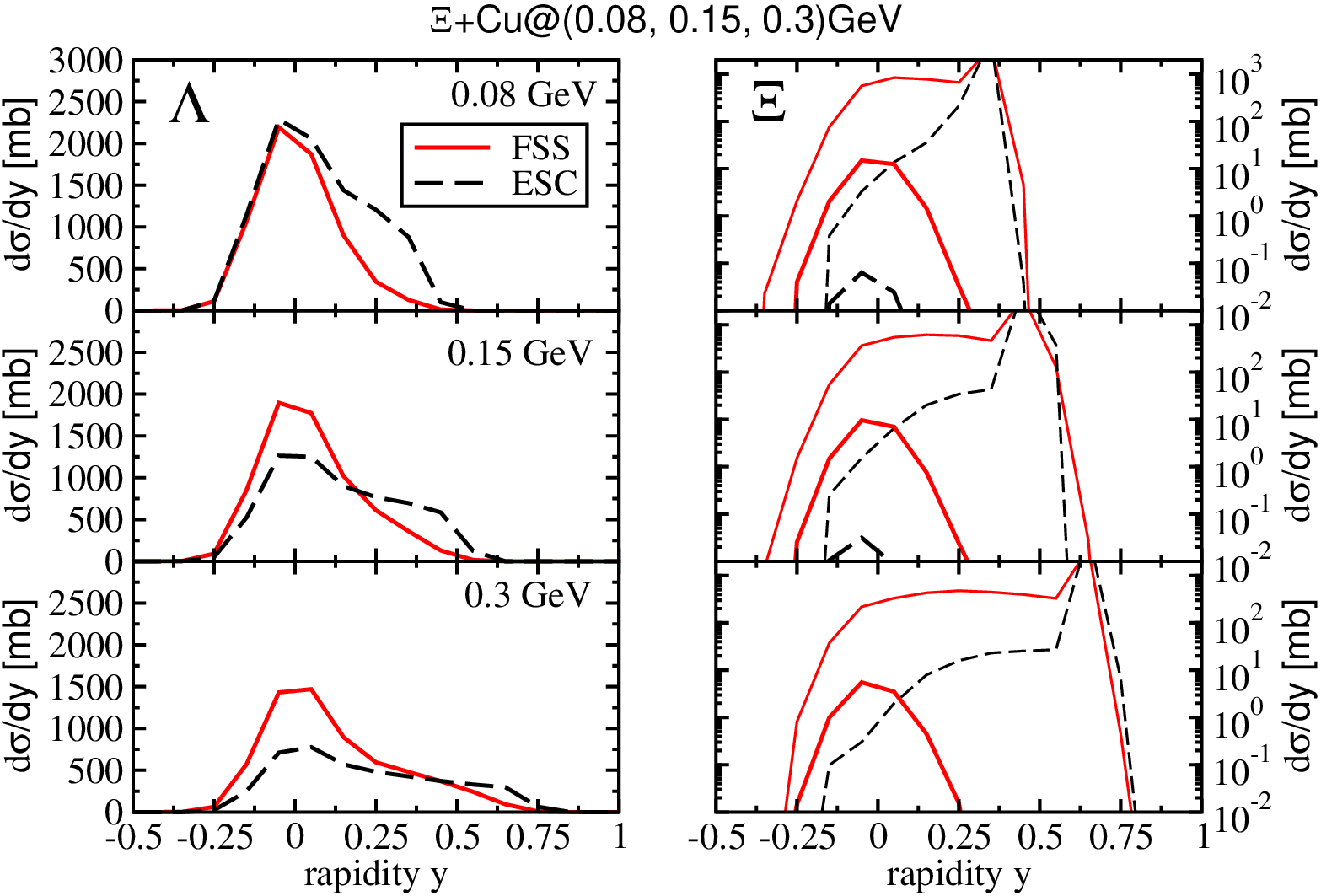}
%\vspace{-1.0cm}
\caption{\label{fig4}
Rapidity distributions for $\Lambda$ (panels on the left) and $\Xi$ (panels on the right) 
particles for $\Xi$-induced reaction on Cu-target. Transport calculations using the ESC 
(dashed curves) and FSS (solid curves) parametrizations at three $\Xi$-beam energies 
(as indicated) are shown. The additional thick dashed (thick solid) curves on the right 
panels indicate the rapidity distributions of bound $\Xi$ particles only using the ESC  
(FSS) parametrizations.
\vspace{-0.3cm}
}
\end{center}
\end{figure}
%%%%%%%%%%%%%%%%%%
However, the reaction dynamics strongly depend on the underlying approach for the 
$\Xi$N-interaction. The transport calculations adopting the ESC model (panel on the left 
in Fig.~\ref{fig3}) lead to less dynamical effects and less re-scattering processes between the 
$\Xi$-beam and the target nucleons. In fact, the primary $\Xi N\to \Lambda\Lambda$ binary 
process mainly dominates the dynamics and contributes basically to $\Lambda$ hyperon 
production. These findings are compatible with the ESC-based results in Fig.~\ref{fig1}, where 
the inelastic $\Xi N\to \Lambda\Lambda$ elementary channel strongly dominates among the 
elastic channels. Note that the $\Xi N\to \Lambda\Sigma$ elementary process does not appear 
at all. In particular, the $\Xi N\to \Lambda\Sigma$ cross section in the ESC model 
is very small as compared to the $\Xi N\to \Lambda\Lambda$ channel and in the order of 
a few \mb~only. Therefore, secondary $\Sigma N\to \Lambda N$ re-scattering only moderately 
affect the dynamics of the $\Xi$-nucleus reaction. 

The situation is different in the transport calculations using the FSS approach 
(panel on the right in Fig.~\ref{fig3}). In contrast to the results in the ESC model, 
the $\Xi N\to \Lambda\Lambda$ primary process is not the dominant one. Indeed, as one can see 
in Fig.~\ref{fig3}, also the primary $\Xi N\to \Lambda\Sigma$ channel occurs. It gives the 
major contribution to $\Lambda$ production. This result is also consistent with Fig.~6 of 
Ref.~\cite{fuwi}, where at this energy ($E_{\Xi}=0.15$ \GeV~ kinetic energy in the laboratory 
respectively $P_{lab}=0.646$ \GeV/c) the $\Sigma$ production channel 
opens. This enhances the $\Sigma$ production and thus the secondary processes with 
strangeness exchange, e.g., $\Sigma N\to \Lambda N$. This feature together with the additional 
issue that elastic and inelastic channels in the FSS approach are of the same order, 
enhances the dynamical effects of the $\Xi$ particles inside the target nucleus such that 
the relaxation time of $\Lambda$ production increases. 

%%%%%%%%%%%%%%%%%%%%%%%%%%%%%%%%%%%%%%%%%%%%%%%%%%%%%%%%%%%%%%%%%%%%%%%%%%%%%%%%%%%
\section{Formation of multi-strangeness hypernuclei}
%%%%%%%%%%%%%%%%%%%%%%%%%%%%%%%%%%%%%%%%%%%%%%%%%%%%%%%%%%%%%%%%%%%%%%%%%%%%%%%%%%%

We thus conclude that the underlying $\Xi$N-interaction has important dynamical effects 
in low-energy $\Xi$-induced reactions thus expecting essential observable signals in the 
formation of S=-2 hypernuclei. This underlines the large physics potential of future S=-2 
production experiments at \panda. For this reason we study now the formation of 
double-strangeness hypernuclei in $\Xi$-induced reactions and the impact of the 
$\Xi$N-interaction on the hypernuclear yields. 

The identification of hypernuclei in the transport calculations is performed as in 
detail reported in previous work~\cite{Gait13}. At first, one applies the statistical 
multifragmentation model (SMM)~\cite{smm} to the freeze-out configuration of the residual 
nucleus. This method provide us with stable and cold fragments (SMM-fragments). As a next 
step, a momentum coalescence between those SMM-fragments and hyperons is performed leading to 
capture and providing us finally with hyperfragments. 

For the formation of hypernuclei the momentum spectra of hyperons are crucial. 
It is therefore of great interest to study the model dependencies of the $\Xi$N-interaction on the 
momentum distributions of strangeness production. Fig.~\ref{fig4} shows this in terms of the 
rapidity distributions of $\Lambda$ (panels on the left) and $\Xi$ particles (panels on the 
right) for $\Xi$-induced reactions at three low energies, as indicated. At first, the stopping 
power of $\Lambda$ hyperons increases in the calculations with the FSS interaction 
compared to the ESC model. This is because of the enhanced re-scattering involving 
$\Lambda$ particles by the FSS interaction. Again, the total $\Lambda$ yields do not essentially 
depend on the underlying $\Xi$N-interaction model, but the detailed dynamics does. 

The rapidity distributions of the cascade hyperons (panels on the right in Fig.~\ref{fig4}) turn 
out to be more interesting. The calculations with the FSS parametrizations give clear evidence 
on the appearance of bound $\Xi$-hyperons inside the target nucleus (thick solid curves 
in Fig.~\ref{fig4}), while in the results using the ESC model the probability of having bound 
cascade matter is extremely low (thick dashed curves in Fig.~\ref{fig4}). In our calculations 
we identify bound particles as those 
particles whose radial distance is less than the target ground state radius plus 2\fm. 
Note that these spectra 
are extracted at the final stage of the reaction, e.g., at 60 \fm/c. Thus, particles which are 
still inside the nucleus at large time scales (relative to freeze-out) can be definitely considered 
as bound particles. At capture, the $\Xi$ particles carry an average velocity 
$\bar{\beta}_\Xi = <v/c>\sim 0.11$ which is much smaller than e.g. the average velocity of 
a nucleon at the Fermi-surface, $\bar{\beta}_{N}\sim 0.25$. 
The coalescence volume in phase space is thus defined by $R$ and $\bar{\beta}_\Xi$.
%The corresponding mean value for the velocity modulus of these bound $\Xi$ 
%particles is located in a range between 0.107 and 0.11 (in units of [c]) with increasing 
%centrality. This value of about 0.1~[c] can be thus considered as a criterion in the 
%$\Xi$-velocity space for phase-space coalescence models.}

This new feature, reported here for the first time, is indeed very promising 
for the formation of multi-strangeness hypermatter at \panda (see below). The occurrence of bound 
cascade hyperons is attributed to the small inelastic cross sections in the FSS model. In fact, 
as one can see in Fig.~\ref{fig1} (and also in Fig.~6 of Ref.~\cite{fuwi} for 
$\Xi N\to \Lambda\Sigma$), the inelastic 
$\Xi N\to \Lambda\Lambda$ cross sections are smaller relative to the elastic 
ones. Therefore, the secondary $\Xi$-beam particles can enter deeper into the target matter and 
get captured by re-scattering and decelerating with target constituents. On the other hand, 
the probability of having bound $\Xi$ hyperons in the ESC-based transport calculations 
is obviously extremely low. 

We have analyzed these transport calculations at different $\Xi$-beam Energies in terms 
of (hyper)fragmentation. 
At first, we have found that the mass distributions become broader around the target mass 
region and the fission region around half the initial target mass number $A_{init}$ is 
getting filled with matter as the $\Xi$-beam energy increases. This is a typical feature 
of statistical thermodynamical models like the SMM, where with increasing beam energy the 
excitation of the residual nucleus increases and thus more fragmentation processes 
(fission, de-excitation) open as the available energy increases. 

\begin{figure}[t]
%\vspace{-2cm}
\begin{center}
\includegraphics[clip=true,width=0.85\columnwidth,angle=0.]{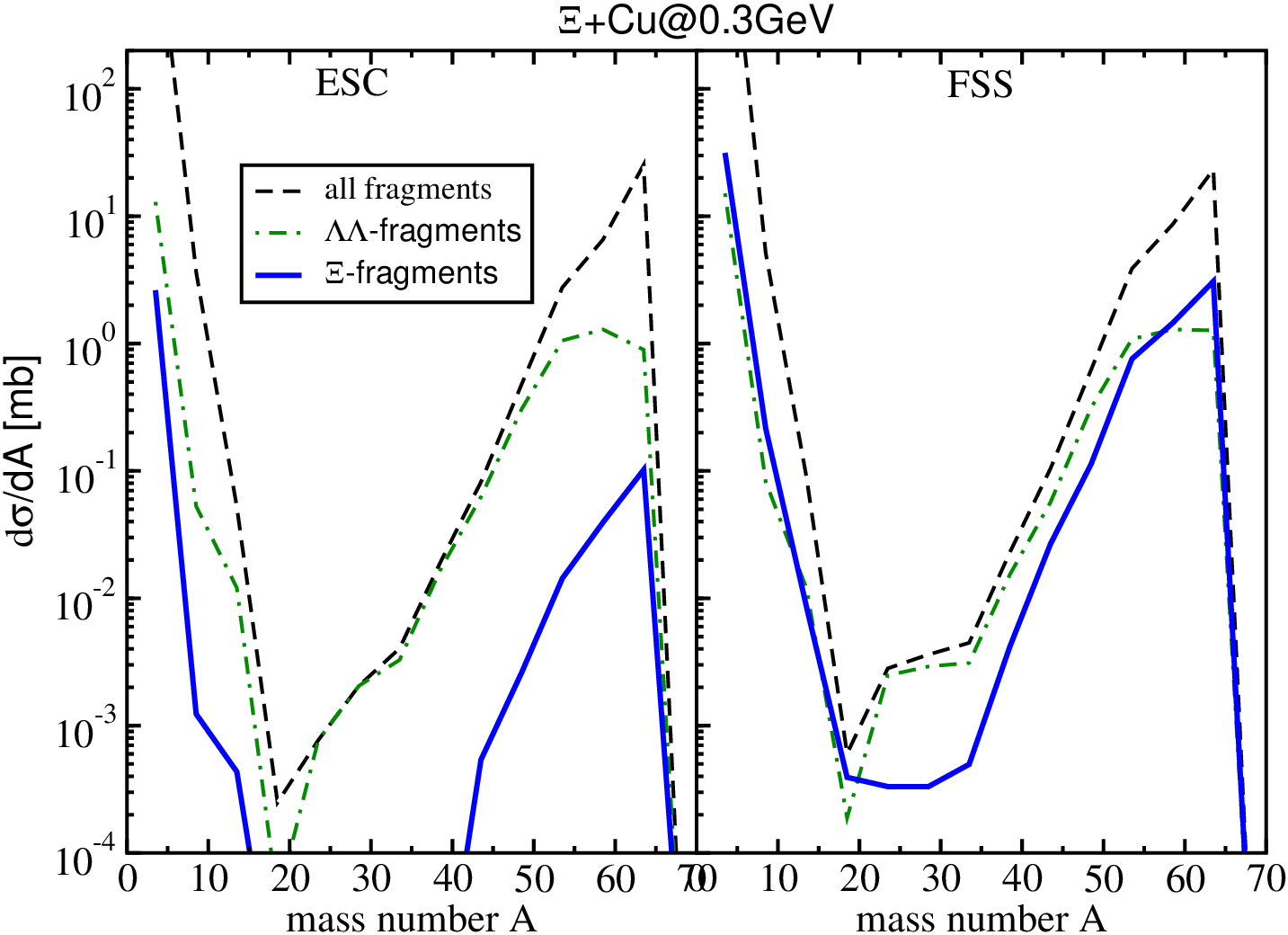}
%\vspace{-1.0cm}
\caption{\label{fig5}
Mass distributions of pure fragments (dashed curves), double-$\Lambda$ fragments (dot-dashed curves) 
and $\Xi$ hyperfragments (thick solid curves). Two sets of calculations are shown, namely one using 
the ESC (panel on the left) and another one using the FSS models (panel on the right). 
The considered reactions are inclusive $\Xi$-induced reactions at a beam $\Xi$-energy of 
of 0.3 \GeV~(respectively $0.937$ \GeV/c~beam $\Xi$-momentum).
\vspace{-0.3cm}
}
\end{center}
\end{figure}

Such an example of fragment distributions is shown in Fig.~\ref{fig5} for SMM-fragments, 
double-$\Lambda$ clusters and $\Xi$-hypernuclei versus their mass number 
for inclusive $\Xi$-induced reactions at a beam energy of $0.3$~\GeV.
It can be seen that both distributions of SMM fragments and double-$\Lambda$ clusters 
become moderately broader in the transport calculations using the FSS parametrizations 
%(panels on the right in Figs.~\ref{fig6a},~\ref{fig6b},~\ref{fig5}) 
(panel on the right in Fig.~\ref{fig5}) 
as compared to the results using the ESC model (panel on the left 
in Fig.~\ref{fig5}). This is due to the increased re-scattering 
processes inside the target nucleus in the FSS-based calculations, as discussed in the previous 
section. This leads to more internal excitation of the residual nucleus and thus to a more 
pronounced fragmentation dynamics leading to increased fragment yields. 

%%%%%%%%%%%%%%%%%%
%%%%%%%%%%%%%%%%%%
The most interesting feature in Fig.~\ref{fig5} is the 
strong dependence of the distributions of $\Xi$-bound hypernuclei on the 
underlying $\Xi$N model. While in the dynamical calculations with the ESC model the production 
probability of bound $\Xi$-matter is relatively low, the calculations using the FSS 
interactions enhances the production of multi-strange $\Xi$-systems considerably. Another 
result of great interest is the prediction of $\Xi$-hypernuclei also far away from the 
regions of ordinary nuclei, e.g., evaporation ($A\sim A_{init}$) and multifragmentation 
($A\leq 4$). This can be clearly seen in Fig.~\ref{fig5}, where the fission region 
($A\sim A_{init}/2$) is filled up by $\Xi$-hypermatter in the calculations with the FSS model. 
This region remains free of multi-strangeness $\Xi$-hypersystems in the ESC-based 
transport calculations.

The differences are of such a magnitude that a sizeable spread is predicted for the S=-2 
production cross sections. It would be therefore a challenge to measure exotic 
multi-strangeness hypermatter at \panda in the future, in order to better
constrain the experimentally still unknown and theoretically very controversial $\Xi$N-interaction, 
eventually ruling out certain approaches to YN and YY interactions.

%%%%%%%%%%%%%%%%%%%%%%%%%%%%%%%%%%%%%%%%%%%%%%%%%%%%%%%%%%%%%%%%%%%%%%%%%%%%%%%%%%%
\section{\label{sec4}Summary and conclusions}
%%%%%%%%%%%%%%%%%%%%%%%%%%%%%%%%%%%%%%%%%%%%%%%%%%%%%%%%%%%%%%%%%%%%%%%%%%%%%%%%%%%

In summary, we have extended our previous studies on hypernuclear physics by the 
consideration of more recent models for the high strangeness S=-2 sector and their possible 
influences on observables in reactions relevant for FAIR. For this purpose we have 
studied first the differences between two well-established models for the $\Xi$N-interaction 
and then parametrized their results. Applications of $\Xi$N-interaction models in the dynamics 
of hadron-induced reactions are discussed in detail for the first time. 

We found important dynamical effects for reactions at \panda, depending essentially on the 
underlying $\Xi$N-approach. Strong inelastic (absorption) effects in the elementary scattering 
channels lead to less dynamical effects in $\Xi$-induced reactions, while in the opposite 
case of less pronounced inelasticity the in-medium dynamics is enhanced. As a consequence, 
bound cascade particles occur in hadron-induced reactions in the case of an attractive 
$\Xi$N-interaction model. 
%as one might have expected.

A coalescence analysis is performed to study the production of (multi-)strangeness hypernuclei 
in low energy $\Xi$-induced reactions using two models for the $\Xi$N-interaction. It is 
found that the distributions of pure fragments and double-$\Lambda$ hyperclusters depend 
only moderately on the applied $\Xi$N-model. However, the role of the $\Xi$N-interaction 
is found to be essential in the formation of multi-strangeness $\Xi$-hypernuclei. In fact, 
models which predict an attractive $\Xi$N interaction on a microscopic level, lead also 
to a copious production of $\Xi$-bound hypermatter at \panda, while in the opposite case 
the formation of such multi-strangeness systems is a rare process. In particular, we predict 
the formation of possible exotic multi-strangeness hypermatter in $\Xi$-induced reactions. 
However, there are still remaining uncertainties on the $\Xi$ mean-field dynamics, which we 
have not studied separately. We consider the investigations presented here as pilot studies 
serving to constrain the production of S=-2 hypernuclei. 

We thus conclude the challenge of the future activities at FAIR to understand deeper the 
still little known high strangeness sector of the hadronic equation of state. Note that 
the strangeness sector of the baryonic equation of state is crucial for our knowledge 
in nuclear and hadron physics and astrophysics. 
For instance, hyperons in nuclei do not experience Pauli blocking within the Fermi-sea of 
nucleons. Thus they are well suited for explorations of single-particle dynamics. 
In highly compressed matter in neutron stars the formation of particles 
with strangeness degree of freedom is energetically allowed. Of particular 
interest are hereby the $\Lambda$-,$\Sigma$-, $\Xi$- and $\Omega$-hyperons 
with strangeness S=-1,-2 and -3, respectively. As shown in recent 
studies~\cite{NeutronStars1,Weissenborn:2011kb}, 
these hyperons modify the stiffness of the baryonic EoS at high densities considerably leading 
to the puzzling disagreement with recent observations of neutron stars in the range of 2 solar 
masses.

%%%%%%%%%%%%%%%%%%%%%%%%%%%%%%%%%%%%%%%%%%%%%%%%%%%%%%%%%%%%%%%%%%%%%%%%%%%%%%%%
%%%%%%%%%%%%%%%%%%%%%%%%%%%%%%%%%%%%%%%%%%%%%%%%%%%%%%%%%%%%%%%%%%%%%%%%%%%%%%%%
\section*{Acknowledgments}
This work was supported by BMBF contract 05P12RGFTE, DFG contract Le 439/8, HIC for FAIR, 
and GSI-JLU Giessen collaboration agreement.

%%%%%%%%%%%%%%%%%%%%%%%%%%%%%%%%%%%%%%%%%%%%%%%%%%%%%%%%%%%%%%%%%%%%%%%%%%%%%%%%
%%%%%%%%%%%%%%%%%%%%%%%%%%%%%%%%%%%%%%%%%%%%%%%%%%%%%%%%%%%%%%%%%%%%%%%%%%%%%%%%

\section*{References}

%\begin{thebibliography}{10}

\end{document}